\documentstyle[aps,epsf,multicol]{revtex}
\begin{document}
\draft
\title{New Mechanism for Electronic Energy Relaxation in Nanocrystals}
\author{Ho-Soon Yang, Michael R. Geller, and W. M. Dennis}
\address{Department of Physics and Astronomy, University of Georgia, Athens, 
Georgia 30602-2451}
\date{May 28, 2000}
\maketitle

\begin{abstract}
The low-frequency vibrational spectrum of an isolated nanometer-scale 
solid differs dramatically from that of a bulk crystal, causing the 
decay of a localized electronic state by phonon emission to be inhibited. 
We show, however, that an electron can also interact with the rigid 
{\it translational} motion of a nanocrystal. The form of the coupling is 
dictated by the equivalence principle and is independent of the ordinary 
electron-phonon interaction. We calculate the rate of nonradiative energy 
relaxation provided by this mechanism and establish its experimental 
observability.
\end{abstract}

\vskip 0.10in

 
The dominant mechanism for the low-energy decay of a localized electronic 
impurity state in a macroscopic semiconductor or insulator is usually phonon 
emission \cite{Henderson and Imbusch,Kuzmany}. In an isolated nanometer-scale 
crystal, however, the reduced dimensionality causes a suppression of the 
vibrational density-of-states (DOS) at low energies\cite{Gaponenko}. In 
particular, in a spherical nanoparticle of diameter $d$ there will be an 
acoustic phonon energy gap $\Delta \omega$ of the order of $2 \pi v/d$, where 
$v$ is a characteristic sound velocity of the bulk crystal. The one-phonon 
energy relaxation rate of an electron in the excited state of a two-level system
with energy spacing $\Delta \epsilon$ therefore vanishes (or is greatly 
suppressed) when $\Delta \epsilon$  is less than $\Delta \omega$ 
\cite{relaxation footnote}. Indeed, a dramatic suppression of the phonon-induced
exciton dephasing rate \cite{Misawa etal CPL,Misawa etal JCP} and nonradiative 
relaxation rate \cite{Woggon etal,Yang etal,Yang etal DPC} has been observed 
in nanocrystalline systems.

Nanoparticles are usually coupled to an environment consisting of other 
nanoparticles, a glass or polymer support matrix, or a solid substrate, and 
this mechanical interaction can modify a nanoparticle's low-frequency 
vibrational spectrum\cite{Alivisatos}. Fig.~\ref{TEM image} shows a 
transmission-electron-microscope image of a cluster of 15 nm ${\rm Y_2 O_3}$ 
nanoparticles prepared by gas-phase condensation. Although little is known in 
detail about the effects of environmental interaction on electronic energy 
relaxation in nanoparticles\cite{exception}, it is clear that for the case
of a nanoparticle only {\it weakly} coupled to other nanoparticles or to a 
substrate, phonon emission will still be prohibited because the low-frequency 
modes introduced by interaction with the environment will involve mostly {\it 
rigid} center-of-mass (CM) motion of the nanoparticles, which produces no 
strain. For example, Fig.~\ref{DOS calculation} shows the collective-mode DOS 
for a model (illustrated in the inset) of the nanoparticle cluster of 
Fig.~\ref{TEM image}. Although the cluster possesses modes at frequencies less 
than $\Delta \omega$, which is about $10 \ {\rm cm}^{-1}$ for these 
nanoparticles, these modes cannot couple to an 
impurity state through ordinary electron-phonon interaction.

In this paper we propose a new nonradiative relaxation mechanism caused by the 
{\it inertial} coupling of an electron to the nanoparticle's translational CM 
motion. This interaction is present because an electron bound to an impurity 
center in an oscillating nanoparticle is in an accelerating reference frame, 
and, in accordance with Einstein's equivalence principle, it feels a fictitious 
time-dependent force. We shall show below that this relaxation mechanism is 
operative even at zero temperature, owing to the fact that quantum zero-point 
motion of the CM is sufficient to produce the fictitious force.

In what follows we shall analyze the simplest nanoparticle model displaying the 
new effect; more sophisticated models, as well as the influence of inertial 
coupling on other dynamical processes such as dephasing, are subjects of 
on-going investigations and will be discussed in future publications. The 
present model consists of a single nanoparticle of mass $M$ connected to a
bulk substrate by a few atomic bonds. The effect of the substrate is to subject 
the nanoparticle to a one-dimensional harmonic oscillator potential 
$V = {1 \over 2} M \Omega^2 X^2$ with frequency $\Omega,$ the $X$ direction 
being perpendicular to the plane of the substrate. However, because the CM 
motion is that of a macroscopic harmonic oscillator interacting with many other 
degrees of freedom, such as the phonons of the bulk substrate, it is necessary 
on physical grounds to include energy dissipation (friction) of that oscillator.
Possible rotational motion of the nanoparticle is not important here and will 
be ignored \cite{rotation footnote}. 

Denoting the CM of the nanoparticle by ${\bf R}$, a Hamiltonian in the CM frame 
can be obtained by rewriting the time-dependent Schr\"odinger equation in terms 
of new coordinates ${\bf r}' = {\bf r} - {\bf R}$ and $t' = t$. After a series 
of gauge transformations we obtain
\begin{equation}
H_{\rm CM}=\sum_\alpha \epsilon_\alpha \ \! c_\alpha^\dagger c_\alpha + \sum_n 
\omega_n a_n^\dagger a_n + \Omega b^\dagger b + \sum_{n \alpha \alpha'} 
g_{n \alpha \alpha'} c_\alpha^\dagger c_{\alpha'} (a_n \! + \! a_n^\dagger) 
- g \sum_{\alpha \alpha'} x_{\alpha \alpha'} c_\alpha^\dagger c_{\alpha'} 
(b \! + \! b^\dagger) + \Delta H.
\label{hamiltonian}
\end{equation}
The first term in Eqn.~(\ref{hamiltonian}) is the Hamiltonian for a 
noninteracting two-level system; the other electronic levels can be 
neglected with no loss 
of generality. Here $\epsilon_\alpha$ (with $\alpha = 1,2$) are the energy 
levels of the localized impurity state, $c_\alpha^\dagger$ and $c_\alpha$ are 
electron creation and annihilation operators \cite{spin footnote}. The second 
term in Eqn.~(\ref{hamiltonian}) describes the nanoparticle's {\it internal} 
vibrational dynamics. The $\omega_n$ are the frequencies of the internal modes, 
and the $a_n^\dagger$ and $a_n$ are the corresponding phonon creation and 
annihilation operators. For the case of a perfectly spherical nanoparticle the 
vibrational eigenmodes and eigenvalues can be obtained analytically within 
continuum elasticity theory\cite{Lamb,Takagahara}; the frequency of the lowest 
internal mode (a 5-fold degenerate torsional mode) is approximately 
$2 \pi v_{\rm t}/d$, where $v_{\rm t}$ is the bulk transverse sound velocity 
and $d$ 
is the diameter. These internal vibrational modes have been observed by 
low-frequency Raman scattering \cite{Duval etal,Champagnon etal,Tanaka 
etal,Krauss etal,Krauss and Wise} and by femptosecond pump-probe 
spectroscopy\cite{Cerullo etal}.

The third term in Eqn.~(\ref{hamiltonian}) describes the harmonic dynamics of 
the CM. As discussed above, the nanoparticle is assumed to be constrained to 
move in the $x$ direction only. Hence, the CM translational motion is described 
by a single bosonic degree-of-freedom,
\begin{equation}
b \equiv \sqrt{M \Omega \over 2} \bigg(X + {i \over M \Omega} \ \! P \bigg),
\label{boson}
\end{equation}
where $X$ and $P$ are the $x$-components of the CM position and  momentum. 

The fourth term in $H_{\rm CM}$ is the ordinary leading-order interaction 
between the 
two-level system and the internal vibrational modes. Here $g_{n \alpha \alpha'}$
is the electron-phonon coupling constant; it depends on the detailed 
microscopic structure of the nanoparticle, the nature and position of the 
impurity, and the spatial dependence of the internal vibrational modes. In the
regime of interest here, where phonon emission is inhibited, this 
electron-phonon interaction term can be ignored, and the remaining Hamiltonian 
is that of a two-level atom in a cavity with a single damped mode \cite{Scully 
and Zubairy}.

The fifth term in Eqn.~(\ref{hamiltonian}), which describes the inertial 
coupling between the two-level system and the CM motion, is the focus of the 
present work. Here $x_{\alpha \alpha'} \equiv \langle \phi_\alpha |x| 
\phi_{\alpha'} \rangle$ are dipole-moment matrix elements, which, of course, 
depend on the form of the impurity states $\phi_\alpha({\bf r})$, 
and
\begin{equation}
g \equiv \sqrt{ m^2 \Omega^3 \over 2 M}
\label{coupling}
\end{equation}
is a coupling constant that depends only on the (bare) electron mass $m$ and on 
macroscopic properties of the nanoparticle. The presence of this fifth term can 
be understood as follows: The Hamiltonian for the noninteracting two-level 
system, the first term in $H_{\rm CM}$, is written in a coordinate system 
moving with 
the oscillating nanoparticle, which is a {\it noninertial} reference frame. 
According to the equivalence principle \cite{Einstein}, the electron therefore 
sees an additional uniform force
\begin{equation}
{\bf F}_{\! \rm eff} = - m {\ddot {\bf R}},
\label{effective force}
\end{equation}
where, as before, ${\bf R}$ is the nanoparticle CM. For the case of harmonic 
motion, Eqn.~(\ref{effective force}) can be written as ${\bf F}_{\! \rm eff} = 
m \Omega^2 {\bf R}$. Thus, the potential energy of an electron at position 
${\bf r}$ in the CM frame is
\begin{equation}
U = -m \Omega^2 {\bf R} \cdot {\bf r},
\label{potential}
\end{equation}
which, in one dimension, is equivalent to the fifth term in 
Eqn.~(\ref{hamiltonian}). Although spin indices have been suppressed in 
Eqn.~(\ref{hamiltonian}), it should be understood that the inertial coupling 
between the electron and the CM motion conserves spin.

The final term in Eqn.~(\ref{hamiltonian}), denoted by $\Delta H$, describes an 
interaction between the nanoparticle's translational motion and a bath of other 
harmonic oscillators, such as the phonons of the bulk substrate or the CM 
degrees-of-freedom of other nanoparticles\cite{interaction footnote}. The 
effect of $\Delta H$ is to dissipate energy from the oscillating nanoparticle. 
In the absence of this damping, the inertial coupling causes energy to be 
continuously exchanged between the two-level system and the CM oscillator in a 
Rabi-like fashion; this interesting dissipation-free limit, although not 
relevant for the nanoparticle systems considered here, will be discussed in 
detail elsewhere.

To study the effect of the fifth term in Eqn.~(\ref{hamiltonian}) on the 
electronic energy relaxation rate we calculate the electron self-energy 
perturbatively\cite{resonance footnote}. We assume that $\Delta \epsilon$ is 
sufficiently smaller than $\Delta \omega$ so that ordinary phonon emission is 
prohibited; this allows us to set $g_{n \alpha \alpha'} = 0$. The leading-order 
self-energy is
\begin{equation}
\Sigma (\alpha, i\omega) = -\frac{g^2}{\beta}\sum_{\omega_{\rm B}} 
\sum_{\alpha'} |x_{\alpha \alpha '}|^2 \, G_0(\alpha', i\omega - 
i\omega_{\rm B}) D(i\omega_{\rm B}),
\label{self energy}
\end{equation}
where $\beta$ is the inverse temperature, $\omega_{\rm B}$ is a bosonic 
Matsubara frequency, $G_0(\alpha ,i\omega) = (i \omega - \epsilon_\alpha)^{-1}$ 
is the noninteracting electron Green's function, and $D(i\omega_{\rm B})$ is the
Fourier transform of a phonon propagator 
\begin{equation}
D(\tau) \equiv - \langle T[b(\tau) + \bar{b} (\tau)][b(0) + \bar{b} (0)]\rangle
\end{equation}
that has been renormalized to include the effects of $\Delta H$. The precise 
form of $D(i\omega_{\rm B})$ depends, of course, on $\Delta H$, and in the 
absence of a reliable microscopic model for $\Delta H$ we shall use a (retarded)
 propagator of the form
\begin{equation}
D^R(\omega ) = \frac{1}{\omega - \Omega + i\gamma} - \frac{1}{\omega + \Omega 
+ i\gamma},
\label{renormalized propagator}
\end{equation}
which has a Lorentzian spectral function of width $\gamma$.

The relaxation rate $\tau^{-1} \equiv -2 \ \! {\rm Im} \ \! 
\Sigma^R(\alpha,\epsilon_\alpha) \big|_{\alpha = 2}$ of the excited state at 
zero temperature is found to be
\begin{equation}
\tau^{-1} = 2\pi g^2|x_{12}|^2  \, f \big(\Delta \epsilon  - \Omega \big),
\label{relaxation rate}
\end{equation}
where $f(\omega) \equiv \gamma /\pi(\omega ^2 + \gamma ^2)$ is a Lorentzian 
function of width $\gamma$. 
The fact that relaxation occurs even at zero temperature, when the nanoparticle 
CM is in its ground state, shows that zero-point CM motion is sufficient to
produce a fictitious force.

When the energy separation $\Delta \epsilon$ 
between the two levels is resonant with the frequency $\Omega$ of the 
translational mode, the electronic relaxation rate becomes
\begin{equation}
\tau^{-1}_{\rm res} = 4 g^2|x_{12}|^2  \, \tau_{\rm cm},
\label{resonant rate}
\end{equation}
where $\tau_{\rm cm} \equiv 1/2\gamma$ is the lifetime of the translational 
mode. This can be written (reinstating factors of $\hbar$) as
\begin{equation}
\tau^{-1}_{\rm res} \, \approx \, 1.6 \times 10^{-21} \ {Q \, \Omega_0^2 \, X^2 
\over M} \ s^{-1},
\end{equation}
where  $Q \equiv \tau_{\rm cm} \Omega$ is the quality factor of the CM 
oscillator, $\Omega_0$ is the CM oscillation frequency in measured in 
wavenumbers, $X$ is the dipole moment measured in Bohr radii, and
$M$ is the nanoparticle mass measured in grams.

We believe that it should be possible to observe this novel relaxation 
mechanism by means of an infrared quantum counter experiment\cite{bloembergen} 
in LaF$_3$:Ho$^{3+}$ nanoparticles. The Ho$^{3+}$  $^5$I$_8$(II) level is
4.5~cm$^{-1}$ above the ground state and can be excited using a pulsed far
infrared source. The population of the $^5$I$_8$(II) state can be probed by a 
pulsed visible laser tuned to the $^5$I$_8$(II) $\rightarrow$ $^5$S$_2$(I) 
transition; a measurement of the intensity of the $^5$S$_2$(I) emission as a 
function of the delay between the far infrared and optical pulses would enable 
the relaxation of the $^5$I$_8$(II) to be observed. The resonant relaxation 
rate for a LaF$_3$ nanoparticle with 
10nm diameter is approximately
\begin{equation}
\tau^{-1}_{\rm res} \approx  1.0 \times 10^{-2} \ {\rm s}^{-1} 
\ \ \ \ \ \ \ \ \ \ {\rm for} \ Q = 10^{2}
\end{equation}
and
\begin{equation} 
\tau^{-1}_{\rm res} \approx  1.0 \times 10^{2} \ {\rm s}^{-1} 
\ \ \ \ \ \ \ \ \ \ \ \! {\rm for} \ Q = 10^{6}.
\end{equation}
The second estimate applies to a very weakly damped nanoparticle. It is simple 
to show that, for this nanoparticle, the maximum $Q$ factor allowed by the 
perturbative analysis leading to Eqn.~(\ref{relaxation rate}) is about $10^8$ 
(see footnote \cite{resonance footnote}). Although small, these rates are still
much larger than the radiative rate between these closely spaced levels.

In conclusion, our analysis shows that fictitious forces produce a coupling 
between an impurity state in a doped nanoparticle and the rigid-body CM 
motion. If the CM oscillation 
frequency is near (on the scale of $\gamma$) to the level separation $\Delta 
\epsilon$, this effect provides a mechanism for energy relaxation even when 
conventional phonon emission is prohibited. We find that on resonance the
relaxation rate is proportional to the translational mode lifetime 
$\tau_{\rm cm}$: This means that relaxation is faster when the translational 
mode is weakly coupled to its environment, whereas coupling to an 
appreciably damped CM mode is less effective. Although we do not believe that 
the mechanism proposed here is responsible for the energy relaxation observed 
in Refs.~\cite{Yang etal} and \cite{Yang etal DPC}, our estimates clearly show 
that this somewhat exotic phenomena is experimentally accessible.

This work was supported by a Research Innovation Award from the Research 
Corporation and by National Science Foundation Grant No.~DMR-9871864. 
It is a pleasure to thank Uwe Happek, David Huber, Coates Johnson, Vadim 
Markel, Richard Meltzer, and William Yen for useful discussions.

\begin{figure}
\caption{Cluster of ${\rm Y}_2 {\rm O}_3$ nanoparticles. The mean particle 
diameter is approximately 15 nm.}
\label{TEM image}
\end{figure}
                                      
\begin{figure}
\caption{Histogram of the number of vibrational modes per unit frequency of a 
model nanoparticle cluster, shown in the inset. The frequency is given in 
wavenumbers. The cluster contains about 1300 nanoparticles of identical mass 
$M,$ and between each adjoining nanoparticle is a harmonic spring with a 
stiffness $k$ chosen such that $\sqrt{k/M}$ is equal to $1 \, {\rm cm^{-1}}$.}
\label{DOS calculation}
\end{figure}


                                        

\begin{references}
\bibitem{relaxation footnote} This statement is correct, of course, only when 
there is some intrinsic or extrinsic phonon broadening mechanism that makes 
Fermi's golden rule applicable.
\bibitem{rotation footnote} Rotation would lead to an fictitious time-dependent 
magnetic field acting on the electron, which is less effective at inducing an 
electronic transition.
\bibitem{spin footnote} The electrons are taken to be spinless for simplicity.
\bibitem{interaction footnote} It is implied that $\Delta H$ includes the 
kinetic and potential energies of the bath degrees of freedom as well.
\bibitem{resonance footnote} A rotating-wave analysis shows that perturbation 
theory is valid here when the energy width $\gamma$ of the CM translational 
mode is larger than the interaction energy $g \, |x_{12}|$, as expected by 
dimensional considerations.
\end{references}
\end{document}